\documentclass{elsart}
\usepackage{epsfig}
\usepackage{amsfonts}
\usepackage{amssymb}
\journal{Physica D}
\begin{document}

\begin{frontmatter}
\title{Defect turbulence and generalized statistical mechanics}
\author[KED]{Karen E. Daniels\thanksref{present}}
\thanks[present]{Present address: Department of Physics, Duke Univ., Durham, NC 27708}
\author[CB]{Christian Beck}
\author[EB]{Eberhard Bodenschatz}
\address[KED]{Laboratory of Atomic and Solid State
Physics, Cornell Univ., Ithaca, NY 14853\thanksref{present}}
\address[CB]{School of Mathematical Sciences, Queen Mary, Univ. of London,
Mile End Road, London E1 4NS, UK}
\address[EB]{Laboratory of Atomic and Solid State
Physics, Cornell Univ., Ithaca, NY 14853}
\date{28 February 2003}

\begin{abstract}
We present experimental evidence that the motion of point defects in
thermal convection patterns in an inclined fluid layer is 
well-described
by Tsallis statistics with an entropic index $q \approx 1.5$. The
dynamical properties of the defects (anomalous diffusion, shape of
velocity distributions, power law decay of correlations) are in good
agreement with typical predictions of nonextensive models, over a range
of driving parameters.
\end{abstract}

\end{frontmatter}

\section{Introduction}

In recent years a generalized version of statistical mechanics
proposed by C. Tsallis, dubbed nonextensive statistical
mechanics \cite{tsa1,tsa2,abe}, has been subject to intensive discussion
\cite{cho}. The basic idea underlying this new theoretical approach is
quite simple: suppose that a physical system of sufficient
complexity cannot, for some reason, maximize the usual
Boltzmann-Gibbs-Shannon entropy leading to the usual statistical
mechanics. In such a case, it is reasonable to conjecture that the
system may then maximize some other, more general entropy measure.
In particular, good candidates for such measures are the Tsallis
entropies $S_q$, which depend on a real parameter $q$, the entropic
index. If $q \ne 1$, the
Tsallis entropies are nonextensive, {\it i.e.}~nonadditive for
independent subsystems. For $q=1$, they reduce to the Boltzmann
entropy and the usual statistical mechanics is recovered.

If the Tsallis entropies are maximized subject to suitable
constraints, one arrives at power-law generalizations for the
canonical ensemble, with $q$ being related to the exponent of
the power law. An important (and open) question involves determining
for which types of systems this more general
formalism is physically relevant. We can imagine various
reasons why a particular physical system may not be able to maximize
the usual Boltzmann entropy: long-ranged interactions/correlations,
multifractality, metastability, or simply the fact that the system is
not in equilibrium due to some external forcing. Indeed, recent work
has shown that nonextensive statistical mechanics is particularly
useful in describing
driven nonequilibrium systems. Useful physical applications include
Eulerian \cite{BLS,prl} and Lagrangian \cite{pla,voth} fully-developed
hydrodynamic turbulence, heavy ion collisions
\cite{quarati}, $e^+e^-$ annihilation experiments
\cite{bediaga,e+e-}, as well as economic \cite{borland} and
biological \cite{upad} systems.

Here we report experimental observations of point defect (dislocation)
motion in defect turbulence in inclined layer convection
\cite{daniels1,daniels2}, which are found to be consistent with
Tsallis statistics. To the best of our knowledge, this is the first
observation of Tsallis statistics in a defect turbulent 
pattern-forming system. The defects behave much like
particles obeying a generalized Tsallis-type statistical mechanics,
with an entropic 
index of approximately $q \approx 1.5$. Their dynamical properties
(anomalous diffusion, skewness, power law decay of correlations) are
in good agreement with typical predictions of
nonextensive models over a range of driving parameters.  In addition, 
since this system is but one of many driven systems exhibiting defect 
turbulence, these findings suggest that similar non-extensive behavior
may be found in other pattern-forming systems such as
electroconvection in liquid  
crystals \cite{Rehberg:1989:TWD}, nonlinear optics
\cite{Ramazza:1992:STD}, and auto-catalytic chemical reactions
\cite{Ouyang:1996:TFS}.

\section{Experiment}

\begin{figure}
\centerline{\epsfig{file=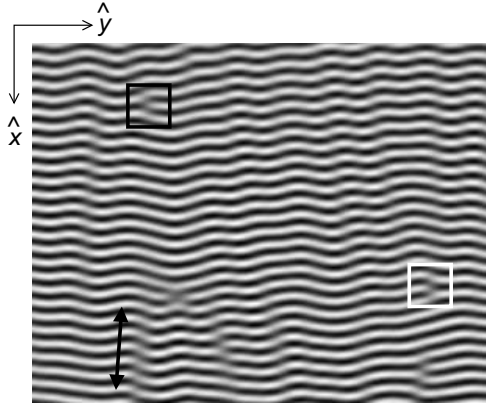,width=2.5in}}
\caption{Example shadowgraph image of undulation chaos
in fluid layer heated from below and cooled from above, inclined by an
angle $\theta = 30^\circ$. The nondimensional driving paramter is
$\epsilon = 0.08$. Black box encloses a positive defect; white, a
negative. Arrow is adjacent to a tearing region of low-amplitude
convection. Uphill direction is at left side of page. Region shown is
the subregion of size $51d \times 63d$ used for analysis.}
\label{convection}
\end{figure}

We examine the motion of defects within spatiotemporally chaotic
patterns formed in experiments on inclined layer convection (ILC), a
variant of Raleigh-B\'enard convection \cite{ABP} in which the thin
fluid layer (heated from below and cooled from above) is inclined
by an angle $\theta$ from the horizontal \cite{daniels1}. For
intermediate angles of inclination ($15^\circ < 
\theta < 75^\circ$ at Prandtl number $\approx 1$), the system exhibits
defect turbulence within a pattern of undulating convection
rolls. This state is known as undulation chaos
\cite{daniels1,daniels2}; an example is shown in
Fig.~\ref{convection}, with defects marked by boxes and a tearing
region marked by an arrow.

The experimental apparatus and data collection for this investigation
have previously been discussed in \cite{daniels1,daniels2}, and only
key points are summarized here. The planform of the convection pattern
was obtained using the shadowgraph technique \cite{deBruyn} and a 
digital camera. The fluid used was compressed CO$_2$ with Prandtl
number ${\mathrm Pr}=\nu/\kappa \approx 1$, where $\nu$ is the
kinematic viscosity and $\kappa$ is the coefficient of thermal
expansion. The convection cell had a thickness of $d = (388 \pm 2)
\mu$m and was of dimension $101d \times 50d$, of which only a central
$51d\times63d$ region was used for analysis. The vertical thermal
diffusion time was $\tau_v = d^2/\kappa = 1.53$ sec. All data was
collected at an inclination of $\theta = 30^\circ$. The strength of
the thermal driving is described by the nondimensional temperature
difference $\epsilon \equiv \frac{\Delta T}{\Delta T_c} -1$; this
quantity was varied in the experiments. Data collection occurred at 3
frames/sec in two modes: four to six hundred 100-frame runs at each of 17
values of $0.04 \le \epsilon \le 0.22$ and a few (4, 2 respectively)
80,000-frame runs at $\epsilon = 0.08$ and $\epsilon = 0.17$. Each
value of $\epsilon$ was reached via quasistatic temperature
changes. We also conducted a sequence of measurements with
quasistatic temperature increases followed by quasistatic temperature
decreases to check for possible 
hysteresis, which was not observed. The Boussinesq parameter
\cite{ABP} was $Q = 0.8$, indicating a breaking of the up-down
($\hat{z}$) symmetry. 

Due to the inclination of the fluid layer, the system is anisotropic
and the convection roll axes align along the direction of the shear flow
($\hat{y}$) for intermediate inclinations such as $\theta =
30^\circ$. In contrast to what is observed in Rayleigh-B\'enard
convection, the defect motion is observed to be primarily transverse
to the convection rolls (``glide,'' in the language of crystals)
instead of parallel (``climb''). As such, the motion of
the defects does not typically adjust the local wavenumber of the
rolls, rather their orientation and the wavenumber of the undulations.

\section{Nonextensive statistical mechanics of defects}

\subsection{Observed defect statistics}

Topological defects in spatiotemporally chaotic patterns such as
undulation chaos in ILC behave much like particles: they have
charge; they are created and annihilated; and they have a
well-defined position and velocity. Their topological charge is a
quantity analogous to a Burger's vector; point defects
(dislocations) are located at positions where there is a
discontinuity of $\pm 2 \pi$ in the phase of the pattern $\Phi$ along
a contour around the point:
\begin{equation}
\oint \vec{\nabla} \Phi \cdot d \vec{s} = \pm 2 \pi,
\end{equation} 
As seen in Fig.~\ref{convection}, undulation chaos
consists of an underlying set of stripes (convection rolls) which
contain both undulations and such defects. Each of these features is
observed to be 
spatiotemporally chaotic. While undulation chaos has strong
nonequilibrium properties due to the thermal forcing, it
nonetheless exhibits stationarity in distributions of the number
of defects and the wavenumber \cite{daniels1}.

If the defects behaved like an ideal gas obeying standard statistical
mechanics, the probability $P_N$ to observe $N$ particles in a
certain area would be given by the Poisson distribution $P_N =
\frac{e^{-\lambda} \lambda^N}{N!}$ where $\lambda$ is the
average number of particles. However, this distribution has been observed 
to be invalid for a number of defect
turbulent systems \cite{Gil,daniels1,Riecke}. A particular model for
$P_N$, based on empirically determined rate equations, has been
studied in \cite{daniels1}. It yields a modified Poisson distribution
\begin{equation}
P_N = \frac{a^{(b/2)+N}}{I_b(2\sqrt{a}) \Gamma (1+b+N)N!}
\label{modpoiss}
\end{equation}
where $I_b$ denotes the modified Bessel
function and the constants $a$ and $b$ are determined by the defect
gain and loss rates. This prediction differs significantly in its
variance from the Poisson distribution, and agrees very well with the
experimentally observed distributions in undulation chaos
\cite{daniels1}. 

Besides defect number, it is also interesting to look at the
statistics of defect velocities. We examine the ensemble of defect
trajectories decomposed into the two directions $\hat{x}$ (across
rolls) and $\hat{y}$ (along rolls, uphill-downhill) coinciding with
the anisotropy of the system. This 
decomposition is done because their motions are fundamentally
different with respect to the underlying roll structure; their
dynamics have been observed to evolve in a nearly statistically
independent way 
\cite{daniels2}. If the defects behaved like ordinary particles
obeying ordinary statistical mechanics, one would expect Gaussian
velocity distributions and normal diffusion of the positions.
Instead, significant deviations are observed \cite{daniels2}.

\begin{figure}
\centerline{\epsfig{file=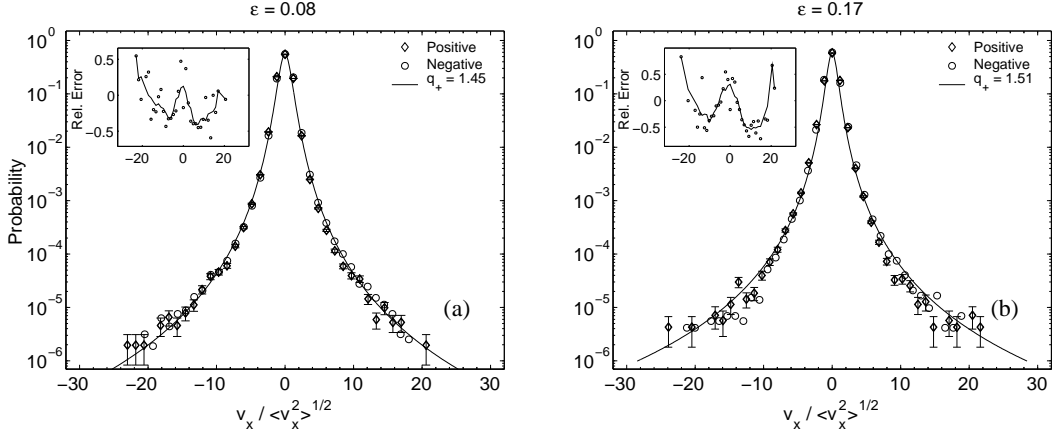,width=\linewidth}}
\caption{Transverse velocity ($v_x$) distributions at (a) $\epsilon =
0.08$ and (b) $\epsilon = 0.17$ for positive and negative
defects, rescaled to $\sigma = 1$. Solid lines are one-parameter fits
to Eq.~(\ref{e:symdist}) for positive defects. Unrescaled
standard deviations were: (a) $\sigma_+ = 0.550$, $\sigma_- = 0.553$
and (b) $\sigma_+ = 0.586$, $\sigma_- = 0.565$. Inset: Relative error
of experiment and theory: $(p_{exp} - p_{theory})/p_{theory}$.}
\label{vxdist}
\end{figure}

\begin{figure}
\centerline{\epsfig{file=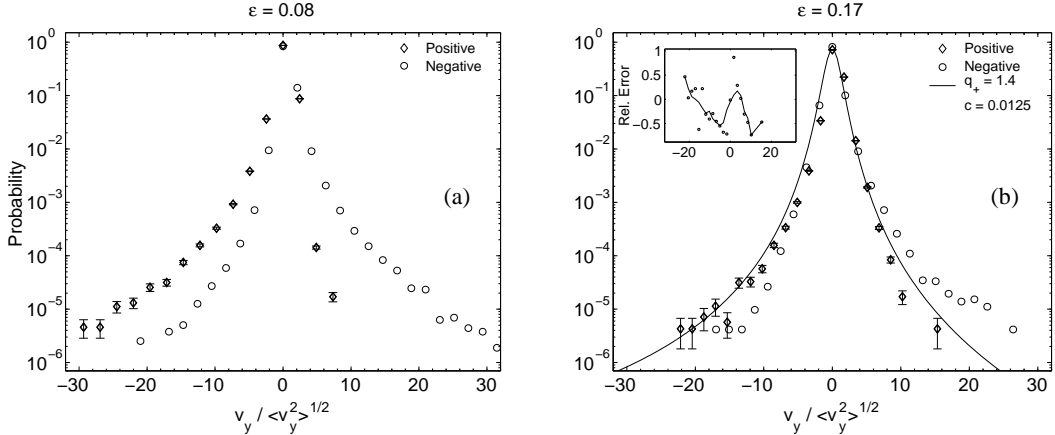,width=\linewidth}}
\caption{Longitudinal velocity ($v_y$) distributions at (a) $\epsilon =
0.08$ and (b) $\epsilon = 0.17$ for positive and negative
defects, rescaled to $\sigma = 1$. Solid line is a
fit to Eq.~(\ref{e:asymdist}) for positive defects. Unrescaled
standard deviations were: (a) $\sigma_+ = 0.099$, $\sigma_- = 0.115$
and (b) $\sigma_+ = 0.141$, $\sigma_- = 0.128$. Inset: Relative error
of experiment and theory: $(p_{exp} - p_{theory})/p_{theory}$.}
\label{vydist}
\end{figure}

The probability density functions (PDFs) of the velocities $v_x$ and
$v_y$ are shown in Fig.~\ref{vxdist} and \ref{vydist}, with velocities
rescaled by their standard deviation $\sigma$. Data is shown for
$\epsilon = 0.08$ and 0.17, the two values with the largest
datasets. The PDFs deviate significantly from a Gaussian, with
velocities of more than $20\sigma$ observed in the long tails. In the
transverse direction ($\hat{x}$), the velocities are several times
faster than those in in the longitudinal ($\hat{y}$) direction.

\begin{figure}[b]
\centerline{\epsfig{file=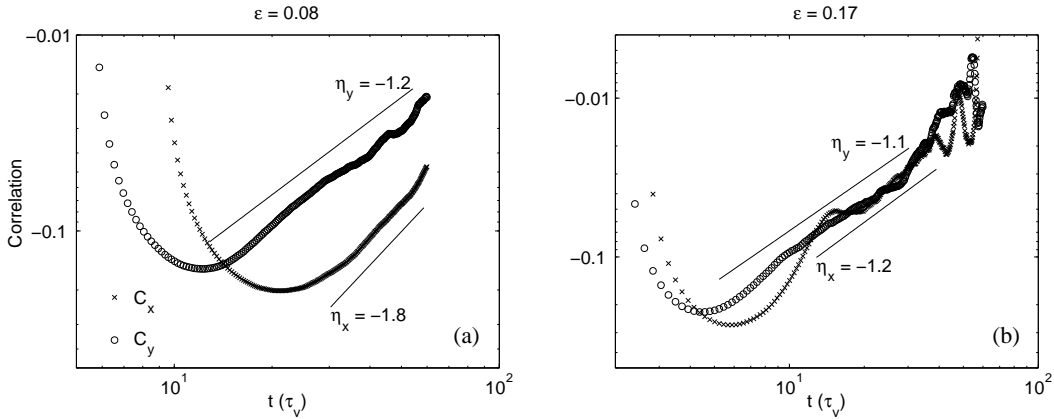, width=\linewidth}}
\caption{Tails of velocity autocorrelation functions for $v_x$ ($\times$)
and $v_y$ ($\circ$) at (a) $\epsilon = 0.08$ and (b) $\epsilon =
0.17$. Straight lines indicate fits to power laws. These curves are
the tails of the results shown in Fig.~4 of \cite{daniels2}.} 
\label{vcorr}
\end{figure}

A common characteristic of nonextensive systems is the
presence of long-range interactions and/or long-range
correlations: for generic nonextensive models, velocity correlation
functions are expected to decay asymptotically with a power law.
While the temporal correlation function $C(t)$ of defect velocities
decays exponentially for small $t$ ($< 10 \tau_v$) \cite{daniels2}, 
Fig.~\ref{vcorr} reveals a superimposed long-term power law
contribution of the form $C(t) \sim t^{-\eta}$ for large 
$t$. For both $v_x$ and $v_y$ at both values of $\epsilon$, the
exponent $\eta$ is greater than 1, although the fit region is of
limited length due to finite observation times. 

\begin{figure}
\centerline{\epsfig{file=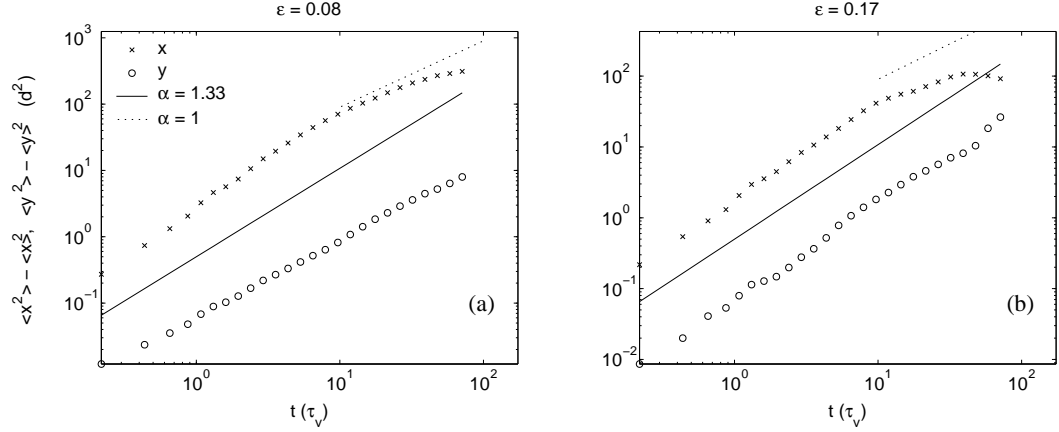,width=\linewidth}}
\caption{Second moment of position trajectories in $\hat{x}$ and
$\hat{y}$. Solid line is diffusive behavior predicted by
Eq.~(\ref{ano}) for $q=1.5$; dotted line is normal diffusive limit
($q=1$).}
\label{diffusion}
\end{figure}

Finally, these observed velocities provide for defect
diffusion. Normal diffusive behavior is given by the second moment of
the position $\langle x(t)^2 \rangle - \langle x(t) \rangle^2  \propto
t^\alpha$ where $\alpha = 1$ and $\langle \cdot \rangle$ represents an
ensemble-average.  Defect motion in undulation chaos exhibits
superdiffusive behavior in both $\hat{x}$ and $\hat{y}$, as shown in
Fig.~\ref{diffusion}, where the exponent $\alpha$ is greater than 1.

\subsection{Generalized canonical distributions}

Given this anomalous behavior, we investigate the proposition that
the defects obey a generalized statistical mechanics of the
Tsallis type and work out the corresponding predictions. Because
the number of defects is not constant but fluctuates, such
fluctuations should properly be embedded into a generalized grand
canonical description. To date, however, no general theory of the
nonextensive grand canonical ensemble is known, and we instead use
the canonical nonextensive model. This theory predicts that the
stationary PDF of an observable $u$ should be given by the
generalized canonical distribution
\begin{equation}
p (u)= \frac{1}{Z_q} (1+(q-1)\tilde{\beta} E(u) )^{-\frac{1}{q-1}}.
\label{e:PDF}
\end{equation}
Here $E(u)$ is an effective energy associated with $u$, and
\begin{equation}
Z_q= \int_{-\infty}^{+\infty} (1+(q-1) \tilde{\beta}
E(u))^{-\frac{1}{q-1}}du
\end{equation}
is a normalization constant. $\tilde{\beta}=1/(kT)$ is a (formal)
inverse temperature variable, which essentially fixes the variance
of the distributions, and $q$ is the entropic index.

The distribution in Eq.~(\ref{e:PDF}) is obtained by maximizing the
Tsallis entropies
\begin{equation}
S_q= \frac{1}{q-1} \left( 1- \sum_i p_i^q \right),
\end{equation}
subject to suitable constraints \cite{tsa1,tsa2}. The $\{p_i\}$
are the probabilities of the microstates $i$ of the system. In the
limit $q \to 1$, the usual Boltzmann distribution is recovered
from Eq.~(\ref{e:PDF}).

The lowest-order model is $E(u)=\frac{1}{2} u^2$, which may be
associated with an effective kinetic energy. For this relation,
one obtains the PDF
\begin{equation}
p(u)=\frac{1}{Z_q}\left(1+\frac{1}{2}(q-1)\tilde{\beta} u^2
\right)^{-\frac{1}{q-1}},
\label{e:symdist}
\end{equation}
where
\begin{equation}
Z_q=\sqrt{\frac{2}{(q-1)\tilde{\beta}}} \cdot
\frac{ \Gamma (\frac{1}{2} ) \Gamma (\frac{1}{q-1}-\frac{1}{2})
}{\Gamma (\frac{1}{q-1} )}.
\end{equation}
The distribution in Eq.~(\ref{e:symdist}) has variance 1 if
$\tilde{\beta}$ is chosen to be
\begin{equation}
\tilde{\beta}=\frac{2}{5-3q}
\label{e:scalebeta}
\end{equation}
yielding a single fit parameter $q$.

In general, however, defect interactions, the local pattern, and other
higher-order correlations of the chaotic forces acting on the defects
will produce small correction terms to $E(u)$. The effect of
such higher-order correlations of chaotic driving forces
has been studied in \cite{hilgers,hydro}. The result of those
considerations is that for distributions which have been rescaled to
$\sigma=1$ the effective energy is of the form
\begin{equation}
E(u)=\frac{1}{2} u^2+c\left( u-\frac{1}{3}u^3\right) +O(c^2).
\label{asy}
\end{equation}
Here $c$ is a small constant related to the skewness of the
distribution and whether or not it is nonzero depends on the
universality class of the chaotic driving forces that are acting.
Eq.~(\ref{asy}) is a perturbative result that is valid for
$u \ll c^{-1}$. Neglecting higher-order corrections of $O(c^2)$ we
obtain the formula
\begin{equation}
p(u)=\frac{1}{Z_q}\left( 1+\tilde{\beta} (q-1) \left( \frac{1}{2} u^2-
c(u-\frac{1}{3}u^3)
 \right) \right)^{-\frac{1}{q-1}}
\label{e:asymdist}
\end{equation}
for the probability distribution.

When this formalism is applied to defects, we associate $u$ with
either $v_x$ or $v_y$, rescaled by $\langle v_{x,y}^2 \rangle
^{1/2}$ to obtain an appropriate dimensionless quantity.
Conceptually, this corresponds to associating the $u^2$ term of
$E(u)$ with an effective ``kinetic energy,'' in the language of
ideal gases. In the case of the transverse motion ($\hat{x}$,
Fig.~\ref{vxdist}) of the defects, the measured PDFs are symmetric
and coincide very well with symmetric ($c=0$) Tsallis
distributions of the form Eq.~(\ref{e:symdist}), while for the
$\hat{y}$ motion (Fig.~\ref{vydist}) significant asymmetries are
present and Eq.~(\ref{asy}) is appropriate with  $c \ne 0$. The
presence of this skewness in the $v_y$ may be associated with the
non-Boussinesqness of the fluid flow, which causes the shear flow to
be asymmetric in $\hat{z}$. Previous studies of inclined layer
convection \cite{daniels1} have revealed similar asymmetries: the
undulations drift in the $\hat{y}$ direction.

We fit the experimentally obtained PDFs to either
Eq.~(\ref{e:symdist}) or (\ref{e:asymdist}) by minimizing $\sum
(p_{exp} - p_{theory})^2 /p_{theory}$, which allows
for appropriate weighting of the tails. These fits can be seen as the
solid lines in Figs.~\ref{vxdist} and \ref{vydist}. For $v_x$, we
obtained values of $q= 
1.45$ at $\epsilon = 0.08$ and $1.51$ at $\epsilon = 0.17$, with the
value for $\tilde{\beta}$ set via Eq.~(\ref{e:scalebeta}) by rescaling
the distributions to $\sigma = 1$.  For $v_y$, we find $q = 1.40$ and
$c = 0.0125$ (such that the expansion is valid for $u \ll  80$) at
$\epsilon = 0.17$. For $v_y$ at $\epsilon = 0.08$, the data is
sufficiently skewed to violate the conditions under which the
expansion in $c$ is valid, and no good fit was found even for values
of $c > 0.03$. It should be noted that these fits do display small
systematic deviations from the data, which may or may not be
significant over six orders of magnitude.

\begin{figure}
\centerline{\epsfig{file=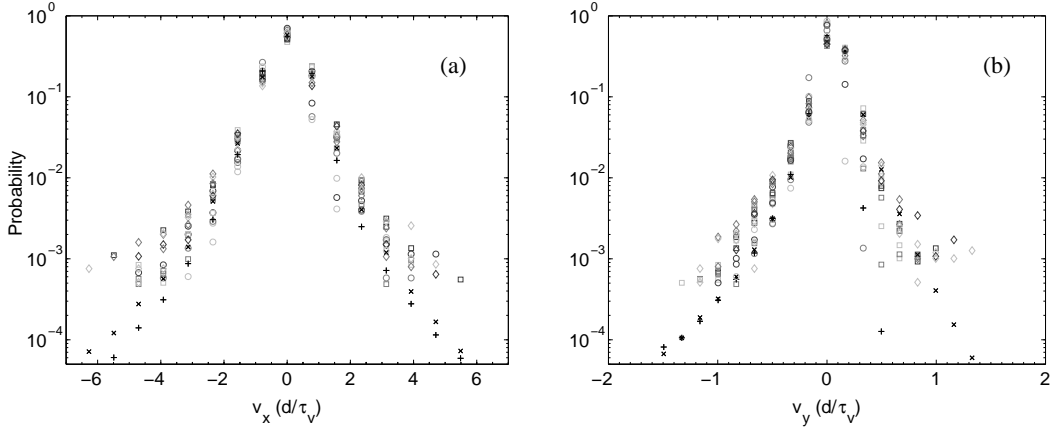,width=\linewidth}} 
\caption{Overlay of (a) $v_x$ PDFs and (b) $v_y$ PDFs for positive
defects at 17 values of $\epsilon$ (open, gray symbols) with data at
$\epsilon = 0.08$ ($+$) and $\epsilon=0.17$ ($\times$). This data
corresponds to the peak region of Figs.~\ref{vxdist} and \ref{vydist},
but not rescaled by $\sigma$.}  
\label{vpdfs}
\end{figure}

\begin{figure}
\centerline{\epsfig{file=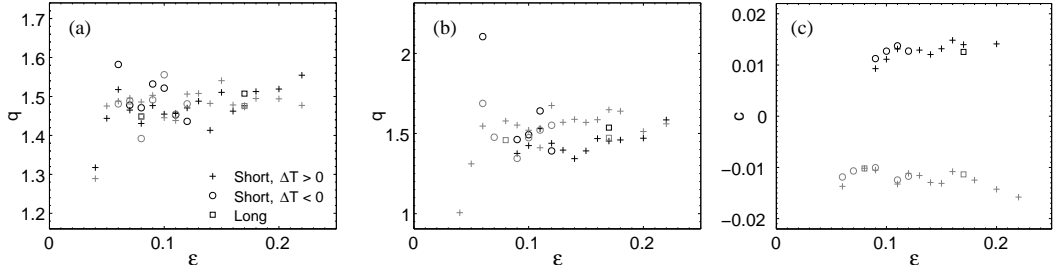,width=\linewidth}}
\caption{Values of fit parameters (a) $q$ for $v_x$ (b) $q$ for
$v_y$, and (c) $c$ for $v_y$ as a function of $\epsilon$. Data in black
is for positive defects; data in gray is for negative. Short runs
in which data was obtained by quasistatic temperature increases from
below are marked with $+$, and the converse with $\circ$.}
\label{epsq}
\end{figure}

The PDFs for the velocity components are not significantly
changing in $q$, $\tilde{\beta}$, and $c$ across the values of
$\epsilon$ surveyed. This is demonstrated in Fig.~\ref{vpdfs} for
$v_x$ and $v_y$ in which the PDFs are found to coincide with no
rescaling. Fig.~\ref{epsq} shows the fit values for $q$ and $c$ at
various $\epsilon$. A mean value of $\bar{q} =  1.48 \pm 0.05$ is
observed for $v_x$. For $v_y$, we observe $\bar{q} = 1.51 \pm
0.15$ and $\bar{c} =  0.012 \pm 0.002$ (with the sign of $c$
corresponding to the sign of the defect). Remarkably, the
distributions are constant in spite of a transition \cite{daniels2}
from disordered undulations for $\epsilon \lesssim 0.1$ to
intermittently ordered undulations for $\epsilon \gtrsim 0.1$.

\subsection{Effective degrees of freedom of the environment}

We now consider concrete dynamical models exhibiting
nonextensive behavior. One of the simplest models, described in
\cite{prl}, is a linear Langevin equation whose friction parameter
is also stochastic. We use such an equation as a simplified
dynamical model for defect velocities:
\begin{equation}
\dot{u} = - \gamma(\vec{x}, t) u + \kappa L(t), \label{Langevin}
\end{equation}
where $L(t)$ is Gaussian white noise of strength $\kappa$ and
$\gamma (\vec{x},t)$ is a random damping constant which fluctuates and may be
attributed to
spatiotemporal changes in the environment of the defect. One can
let $\kappa$ fluctuate as well \cite{prl}, but this is not
necessary for our purposes.

It has been shown in \cite{prl} that Eq.~(\ref{Langevin}) generates
Tsallis distributions for $u$ if $\beta \equiv \gamma/\kappa^2$ is
$\chi^2$-distributed with degree $n$, {\it i.e.}~if the PDF of
$\beta$ is given by
\begin{equation}
f (\beta) = \frac{1}{\Gamma(\frac{n}{2})}
\left( \frac{n}{2\beta_0} \right) ^\frac{n}{2} 
\beta^{\frac{n}{2}-1}
\exp \left[ -\frac{n\beta}{2\beta_0} \right] . 
\label{fluc}
\end{equation}
Such a distribution is generated by a sum of $n$ Gaussian random
variables $X_i$, squared:
\begin{equation}
\beta =\sum_{i=1}^{n} X_i^2 \label{nat}
\end{equation}
If $\beta$ fluctuates on a time scale that is much larger than the
relaxation time to reach local equilibrium, then, as shown in
\cite{prl}, Eq.~(\ref{Langevin}) generates stationary densities of
the form (\ref{e:symdist}) with
\begin{eqnarray}
q=1+\frac{2}{n+1}, \label{qq} \\
\tilde{\beta}=\frac{2}{3-q} \beta_0.
\end{eqnarray}
Here $\beta_0=\int_0^\infty f(\beta) \beta~d\beta$ is the average
value of the fluctuating $\beta$. The experimentally observed
value $q\approx 1.5$ for the defect statistics thus means that
there are effectively 3 independent degrees of freedom that
contribute to the fluctuating local defect environment.

Defects are no ordinary particles: they have neither a well-defined
mass nor a well-defined size. Furthermore, the nature of defect
turbulence means that they are moving within a spatiotemporally
chaotic environment. Therefore, one expects that there is an 
ensemble of damping constants which depend on the environment of the
defect, providing the fluctuating effective damping $\gamma$.

In particular, the fastest velocities result from circumstances in
which the defect is moving in a local environment corresponding to
weak damping (small $\gamma$). In undulation chaos, such events
coincide with the presence 
of a transverse region of low-amplitude convection, as shown by the
black arrow in Fig.~\ref{convection}.  The rapid motion takes place
due to the fact 
that the pattern ``tears'' in such a way that each roll broken
rejoins with the one next to it, rapidly transporting the
defect across the weak region.  The driving
forces $L(t)$ are weakly damped during such a time interval,
leading to very large velocities for short intervals, until
another region with another $\gamma$ is reached. The behavior of
such flight-like events is further described in \cite{daniels2}.

In general, one may also consider probability distributions other
than the $\chi^2$-distribution in used Eq.~(\ref{fluc}). These lead to
statistics other than Tsallis statistics: {\it
superstatistics} \cite{cohen}, for which generalized versions of 
statistical mechanics can be constructed
as well \cite{souza}. Using these extensions, it can been shown
that for $\beta$-fluctuations of small variance and not too large
$|u|$ all superstatistics are universal,  and in fact are given by
Tsallis statistics \cite{cohen}. Significant differences arise only for
large $|u|$. In general, the $\chi^2$-distribution is
distinguished by the fact that it yields a superstatistics whose
generalized entropies have ``better'' properties than those
of other superstatistics. 

\subsection{Anomalous diffusion}

Typical nonextensive models predict anomalous diffusion, such that
\begin{equation}
\langle x(t)^2 \rangle - \langle x(t) \rangle^2\propto t^\alpha
\label{e_diff}
\end{equation}
with $\alpha
\ne  1$, as is observed in the diffusion of defects in undulation
chaos. The nonlinear Fokker-Planck model previously discussed in
\cite{bukman,plastino} describes correlated anomalous diffusion in
the framework of nonextensive models. It is based on a generalized
Fokker-Planck equation of the form
\begin{equation}
\frac{\partial}{\partial t} p(x,t) =
-\frac{\partial}{\partial x} \left( F(x)p(x,t) \right) 
+D\frac{\partial^2}{\partial x^2} (p(x,t))^\nu
\end{equation}
with a linear drift term given by $F(x)=k_1-k_2x$.

The linear case ($\nu=1$) corresponds to the standard Fokker-Planck
equation generating the usual Brownian motion processes with
Gaussian densities. As shown in \cite{bukman}, the nonlinear case where
$\nu \ne 1$ generates Tsallis distributions, with the entropic
index $q$ being related to the parameter $\nu$ by $q=2-\nu$.
In addition, the model generates anomalous diffusion for $\nu \ne 1$,
with the relation between the exponent $\alpha$ of anomalous diffusion
and the entropic index $q$ given by \cite{bukman}
\begin{equation}
\alpha = \frac{2}{1+\nu} = \frac{2}{3-q}.
\label{ano}
\end{equation}

Thus, we can compare the results for the Tsallis fits to the PDFs and
the exponents in the diffusion graphs. Remarkably, the value
$\alpha=1.33$ obtained from Eq.~(\ref{ano}) for $q=1.5$ provides a
good match to the observed data. Other models, based on 2-dimensional
Levy flights \cite{zanette}, predict $\alpha = q-1\approx 0.5$ and are
apparently not consistent with our data. This can be understood within
the framework described here: we earlier found that $n \approx 3$,
which means that there are effectively more than 2 degrees of freedom
contributing to the fluctuating environment of the defect. 

\subsection{Velocity correlation function}

To better understand the relationship between Figs.~\ref{vcorr} and
\ref{diffusion}, we can perform a simplified calculation relating the 
velocities and positions. To do so, we neglect all spatial
correlations, which do effect the diffusion but have not been
measured, and focus only on the temporal ones. In each spatial
direction the displacements obey $x(t) \equiv \int_0^tv(t')dt'$.
Hence,  
\begin{eqnarray}
\langle x^2(t) \rangle &=& \int_0^tdt'\int_0^t dt''\langle
v(t')v(t'')\rangle \\ &=& 2\int_0^tdt' \int_0^{t'}d\tau \langle
v(0)v(\tau ) \rangle. 
\label{cor}
\end{eqnarray}
Assuming the simple asymptotic form $\langle v(0)v(\tau) \rangle
\sim \tau^{-\eta}$, the two integrations in Eq.~(\ref{cor}) yield
\begin{equation}
\langle x^2 (t) \rangle \sim t^{-\eta+2},
\end{equation}
as the expected behavior in the asymptotic limit.
By equating this expression with Eq.~(\ref{e_diff}), one obtains a
relation between 
$\eta$ (describing the velocity autocorrelation) and $\alpha$
(describing the diffusion), namely
\begin{equation}
\eta = 2 - \alpha . 
\label{eta}
\end{equation}
In the experimental data displayed in Fig.~\ref{diffusion}, we observed 
that the effective anomalous diffusion exponent $\alpha$ is given by 
$\alpha \approx 1.33$ on moderate time scales and decreases to $\alpha
\lesssim 1$ for large time scales $t>10 \tau_v$.  
Thus, Eq.~(\ref{eta}) yields  $\eta \approx 2-\alpha
\gtrsim 1$ for the long-term decay exponent of
the velocity correlation function. As Fig.~\ref{vcorr} shows, this is
in good agreement with the measured decay rate of the velocity
autocorrelation function. 

\section{Conclusions}

We have provided experimental evidence that topological defects in 
Inclined Layer Convection behave in a similar fashion to an ideal gas of nonextensive
statistical mechanics with $q \approx 1.5$. This is an application to
a physical system far from equilibrium, where the standard statistical
mechanics has little to say. Not only do the 
measured probability distributions of defect velocities agree well
with Tsallis statistics, but we find quantitative agreement in the
nonextensive parameter $q \approx 1.5$ as determined from both
velocity distributions and diffusive behavior.

While the direct observation of nonextensive behaviour of the entropy
may not be possible , further work may nonetheless be able to provide a
prediction for $q$. One promising approach would be to quantify
the fluctuations in $\beta$, the environment of the defect, in
terms of observed pattern properties such as local wavedirector and
convection amplitude. If so, then the recent superstatistics described in
\cite{cohen} could provide a suitable framework. Furthermore,
other defect-turbulent systems  may also exhibit similar behavior and
prove a fertile ground for further investigations.

KED and EB are grateful to the National Science Foundation for support
under DMR-0072077.


\begin{thebibliography}{99}

\bibitem{tsa1} C. Tsallis, {\it J. Stat. Phys.}, {\bf 52}: 479 (1988)

\bibitem{tsa2} C. Tsallis,  R.S~. Mendes and A.~R. Plastino, {\it
Physica}, {\bf 261A}: 534 (1998)

\bibitem{abe} S. Abe, Y. Okamoto (eds.),
{\it Nonextensive Statistical Mechanics and Its Applications},
Berline: Springer (2001)

\bibitem{cho} A. Cho, {\it Science}, {\bf 297}: 1268 (2002)

\bibitem{BLS} C. Beck, G.~S. Lewis, H.~L. Swinney, {\it Phys. Rev. E},
{\bf 63}: 035303R (2001)

\bibitem{prl} C. Beck, {\it Phys. Rev. Lett.}, {\bf 87}: 180601
(2001)

\bibitem{pla} C. Beck, {\it Phys. Lett. A}, {\bf 287}: 240 (2001)

\bibitem{voth} G.~A. Voth, {\it et al.}, {\it J. Fluid Mech.}, {\bf
469}: 121 (2002)

\bibitem{quarati} W.~M. Alberico, A. Lavagno, P. Quarati, {\it Eur.
Phys. J.}, {\bf C12}: 499 (2000)

\bibitem{bediaga} I. Bediaga, E.~M.~F. Curado, J.~M. de Miranda,
{\it Physica}, {\bf 286A}: 156 (2000)

\bibitem{e+e-} C. Beck, {\it Physica}, {\bf 286A}: 164 (2000)

\bibitem{borland} L. Borland, {\it Phys. Rev. Lett.}, {\bf 89}: 098701 (2002)

\bibitem{upad} A. Upadhyaya, J.-P. Rieu, J.~A. Glazier, Y. Sawada,
{\it Physica}, {\bf 293A}: 549 (2001)

\bibitem{daniels1} K.~E. Daniels, E. Bodenschatz, {\it Phys. Rev. Lett.},
{\bf 88}: 034501 (2002)

\bibitem{daniels2} K.~E. Daniels, E. Bodenschatz, {\it Chaos}, {\bf
13}: 55 (2003)

\bibitem{Rehberg:1989:TWD} I. Rehberg, S. Rasenat, and V. Steinberg,
{\it Phys. Rev. Lett.}, {\bf 62}:  756  (1989).

\bibitem{Ramazza:1992:STD} P. Ramazza, S. Residori, G. Giacomelli, and
F. Arecchi, {\it Europhys. Lett.},   {\bf 19}:  475  (1992).

\bibitem{Ouyang:1996:TFS} Q. Ouyang and J.~M. Flesselles, {\it
Nature}, {\bf 379}:  143  (1996). 

\bibitem{ABP} G.~Ahlers, E. Bodenschatz, W. Pesch, {\it
Annu. Rev. Fluid Mech.}, {\bf 32}: 709 (2000)

\bibitem{deBruyn}J.~R. de~Bruyn {\it et al.}, {\it
Rev. Sci. Instrum.}, {\bf  67}:  2043 (1996)

\bibitem{Gil}L. Gil, J. Lega, J. L. Meunier, {\it Phys. Rev. A}, {\bf
41}: 1138 (1990)

\bibitem{Riecke}Y.-N. Young, H. Riecke, physics/0209062

\bibitem{hilgers} A. Hilgers and C. Beck, {\it Phys. Rev. E}, {\bf 60}:
5385 (1999)

\bibitem{hydro} C. Beck, {\it Physica}, {\bf 277A}: 115 (2000)

\bibitem{cohen} C. Beck, E.~G.~D. Cohen, cond-mat/0205097, to
appear in Physica A

\bibitem{souza} C. Tsallis, A.~M.~C. Souza, {\it Phys. Rev. E}, {\bf
67}: 026106 (2003) 

\bibitem{bukman} C. Tsallis, D.~J. Bukman, {\it Phys. Rev. E}, {\bf 54}:
R2197 (1996)

\bibitem{plastino} A.~R. Plastino, A. Plastino, {\it Physica}, {\bf
222A}: 347 (1995)

\bibitem{zanette} D.~H. Zanette, P.~A. Alemany, {\it Phys. Rev. Lett.},
{\bf 75}: 366 (1995)

\end{thebibliography}
\end{document}